# On the electromagnetic wave interaction with subluminal, luminal, and superluminal mirrors

T. Z. Esirkepov[a] and S. V. Bulanov[a,b]

[a]*Kansai Institute for Photon Science, National institutes for Quantum and Radiological Science and Technology (QST), 8-1-7 Umemidai, Kizugawa, Kyoto 619-0215, Japan*
[b]*Extreme Light Infrastructure ERIC, ELI–Beamlines Facility, Za Radnici 835, Dolni Brezany 25241, Czech Republic*

**Abstract**
As predicted by A. Einstein [Ann. Phys. (Leipzig) 17, 891 (1905)], the electromagnetic wave reflected at a moving mirror is frequency-upshifted and intensified as high as the mirror velocity is close to the speed of light in vacuum. However, at this limit the mirror reflectivity vanishes, because the higher the wave frequency the more transparent matter is. To resolve this paradox, we analyse the electromagnetic wave propagation in medium where the refractive index modulation moves at the speed of light in vacuum. Although the luminal and superluminal modulations are unconditionally transparent for the incident radiation, they both can reflect. We find the new type of the electromagnetic wave reflection with the increasing in time frequency inside the luminal mirror. If the modulation disappears the high frequency radiation is released as a short wavepacket.

## 1. Introduction

According to the famous paper published by A. Einstein in 1905 [1] the electromagnetic wave interaction with a counter-propagating relativistic mirror (in the head-on wave-mirror collision configuration) results in the reflected electromagnetic pulse compression, upshifting of its carrier frequency, and enhancement of its electric field strength. This regime is sometimes called the double Doppler effect. It is of substantial interest for developing compact sources of coherent radiation in the high photon energy range. The reflected pulse frequency and electric field strength are given by the relationships

$$(1) \quad \omega_r = \omega_i\, g_M(\beta_M), \quad E_r = r\, E_i\, g_M(\beta_M),$$

where $\omega_i, E_i$ and $\omega_r, E_r$ are, respectively, the incident and reflected pulse parameters: frequency and amplitude; $\beta_M = v_M/c$ is the normalized mirror velocity; $r$ is the mirror reflectivity. The frequency upshifting factor, playing also the role of the pulse intensification factor, depends on the mirror velocity as

$$(2) \quad g_M(\beta_M) = \left|\frac{1+\beta_M}{1-\beta_M}\right|.$$

In the ultra-relativistic limit, when $\beta_M \to 1$, this factor approximately equals to $g_M(\beta_M) \approx 4\gamma_M^2$, where $\gamma_M$ is the Lorentz factor of the mirror:

$$(3) \quad \gamma_M = 1/\sqrt{1-\beta_M^2}.$$

In medium, the refractive index modulations play a role of semi-transparent mirrors. In general, if such the modulations do not transfer energy and momentum, then they are *phase objects*. In this case their velocity is a phase velocity, which can be subluminal ($\beta_M < 1$), luminal ($\beta_M = 1$), or superluminal ($\beta_M > 1$). As B. M. Bolotovskii and V. L. Ginzburg explained in Ref. [2], the motion with superluminal velocity of the object emitting or reflecting electromagnetic radiation is not forbidden by the special theory of relativity provided this is not the particle velocity or the group velocity of a wave (see also Refs. [3, 4]). When the normalized velocity of the mirror is larger than unity, $\beta_M > 1$, the formula for the frequency upshifting factor is the same as Eq. (2).

The light interaction with moving mirrors has been considered in Refs. [5-8] (see also review article [9] and references therein). Fig. 1 is the Minkowski diagram illustrating the electromagnetic pulse reflection (transmission) at the subluminal and superluminal mirrors. Depending on the mirror reflectivity a portion of the pulse is transmitted through the mirror and another portion is reflected appearing ahead of the subluminal mirror or behind the superluminal mirror. In both these cases the electromagnetic pulse frequency is upshifted according to Eqs. (1) and (2). One can define the factor $g_M$ in Eq. (2) as $g_M = (1 + \beta_M)/(1 - \beta_M)$, including possible change of sign. Then, in the case of subluminal mirror, the reflected frequency is positive as long as the incident frequency is positive. On the contrary, for the superluminal mirror, the reflected frequency is negative. It can be interpreted as the time reversal of the electromagnetic pulse, i.e. the tail of the pulse becomes its head upon reflection [8].

In Refs. [5,6] it was noted that the reflection coefficient vanishes while the reflected frequency infinitely grows when $\beta_M \to 1$. This fact can be elucidated by analyzing the light reflection at the interface between the vacuum region at $x > c\beta_M t$ and plasma region at $x < c\beta_M t$. In the boosted reference frame, where the interface is at rest, we apply the Fresnel formula for the reflection coefficient

$$(4) \quad \mathcal{R}_F = |r|^2 = \left|\frac{1-n}{1+n}\right|^2,$$

with the plasma refractive index $n = \sqrt{1 - \omega_{pe}^2/\omega_{ib}^2}$. Here $\omega_{pe} = \sqrt{4\pi n_e e^2/m_e}$ is the Langmuir frequency (Lorenz invariant); $n_e$ is the electron density; $e$ and $m_e$ are the electron charge and mass, respectively; $\omega_{ib}$ is the incident electromagnetic wave frequency in the boosted reference frame, it is related to the incident radiation frequency $\omega_i$ in the laboratory frame as $\omega_{ib} = \omega_i \sqrt{g_M} \approx 2\gamma_M \omega_i$. When the interface velocity tends to the speed of light in vacuum, $\beta_M \to 1$, the value of the incident frequency in the boosted frame tends to infinity. Therefore, we have $\omega_{pe}^2/\omega_{ib}^2 \ll 1$ and the reflection coefficient is approximately equal to

$$(5) \quad \mathcal{R}_F \approx \frac{\omega_{pe}^4}{16\omega_{ib}^4} = \frac{\omega_{pe}^4}{256\gamma_M^4 \omega_i^4}.$$

We see that the reflection coefficient tends to zero for $\gamma_M \to \infty$, i.e. plasma becomes more and more transparent.

In the model of a thin layer of plasma with the electron density of
$$(6) \quad n_e(x,t) = nl\delta(x - c\beta_M t),$$
where $n$ and $l$ are the layer's electron density and thickness in the laboratory frame of reference, $\delta(\cdot)$ is the Dirac delta function, the reflection coefficient in the boosted frame of reference is (e.g. see Ref. [9])

$$(7) \quad \mathcal{R}_\delta = \frac{\omega_{pe}^4}{\omega_{pe}^4 + \omega_{ib}^2(2c/l)^2} \approx \frac{\omega_{pe}^4 l_0^2}{16\gamma_M^4 \omega_i^2 c^2}.$$

Here $l_0 = \gamma_M l$ is the plasma layer proper thickness in the boosted frame of reference. As in the case of the plasma-vacuum interface, the reflection coefficient vanishes as $\gamma_M^{-4}$ for $\gamma_M \to \infty$.

As we see, the case of a luminal mirror requires special consideration. The tendencies of the reflectivity and the reflected wave frequency at the limit of $\beta_M \to 1$ rise two questions. What is the maximum attainable frequency upshift and the electric field enhancement? Can the luminal mirror reflect? Here we present the investigation on these two questions. Finding the answers on these questions is of crucial importance in view of the relativistic mirror consideration as a mean for reaching critical field of quantum electrodynamics. This requires approaching as close as possible to the limit of $\beta_M \to 1$, using various realizations of relativistic mirrors formed in high-power laser interaction with matter from the plasma waves [10, 11], thin electron layers [12], oscillating mirrors [13, 14] to the accelerated dense plasma slabs [5,6].

Below we will use the terms "superluminal or luminal plasma-vacuum interface or plasma slab". We note that this does not imply the superluminal or luminal motion of plasma particles. These terms merely mean that the local Langmuir frequency, which is essentially the medium response to the electromagnetic field, varies from one point of medium to another with an apparent superluminal or luminal velocity. For example, ionization induced in gases by an external electromagnetic beam rotating like a searchlight can produce a spot of a non-zero Langmuir frequency moving with an arbitrary velocity.

## 2. Maximum attainable electric field enhancement and frequency upshift

As mentioned above, in the limit of $\beta_M \to 1$ the reflectivity tends to zero, while frequency tends to infinity. However, the reflected electric field strength is finite and non-zero. For example, in the case of plasma-vacuum interface, by combining Eqs. (1) and (5) we obtain
$$(8) \quad E_r \approx \mathcal{r} E_i 4\gamma_M^2 \approx (E_i/4)(n_{e0}/n_{cr}).$$
Here $n_{cr} = m_e \omega_i^2/(4\pi e^2)$ is critical electron density corresponding to the incident frequency $\omega_i$, $n_{e0}$ is the electron density of plasma (in the laboratory frame of reference). Although the reflected wave intensity $I_r \sim E_r^2 c/4\pi$ is also finite and non-zero, the reflected energy $\mathcal{E}_r \sim E_r^2 c\tau_r$ tends to zero (but not exponentially) because the reflected wave packet duration $\tau_r \approx \tau/4\gamma_M^2$ is compressed with respect to the incident wave packet duration $\tau$: $\mathcal{E}_r \sim (E_i^2 c\tau/16)(n_{e0}/n_{cr})\gamma_M^{-2}$.

The upper bounds for the frequency upshift and electric field enhancement can be found at the limit of $\beta_M \to 1$ when the reflection is perfect, i.e. $\mathcal{R} = 1$. Consequently, in the boosted reference frame, the incident frequency must be less than the Langmuir frequency and this gives the upper bound on the Lorentz factor:
$$(9) \quad \omega_{ib} \approx 2\gamma_M \omega_i \leq \omega_{pe} \Rightarrow \gamma_M \leq \omega_{pe}/(2\omega_i).$$
Then, in the laboratory reference frame, for the reflected frequency and electric field we have
$$(10) \quad \omega_r \approx 4\gamma_M^2 \omega_i \leq \omega_{pe}^2/\omega_i = (n_{e0}/n_{cr})\omega_i,$$
$$(11) \quad E_r \leq E_i(n_{e0}/n_{cr}).$$

The above logic is almost reversible. Suppose we have $\omega_r = (n_{e0}/n_{cr})\omega_i$. Then, knowing that $\omega_r \approx 4\gamma_M^2 \omega_i$, we obtain $\gamma_M \approx \omega_{pe}/(2\omega_i)$. Therefore, $\omega_{ib} \approx 2\gamma_M \omega_i \approx \omega_{pe}$, which is essentially the condition of an almost total reflection. In other words, the assumption of the perfect reflection is effectively equivalent to the situation when entities at the left-hand side of Eqs. (10) and (11) are equal to their upper bounds. We note that this estimation is especially useful when $n_{e0} > n_{cr}$.

## 3. Electromagnetic wave interaction with moving plasma-vacuum interface

We consider a one-dimensional equation of electromagnetic wave interacting with a moving plasma-vacuum interface:

$$(12) \quad A_{tt} - A_{xx} + n\,\rho(x - \beta t)A = 0.$$

Here $A(t,x)$ is a transverse component of vector-potential; $n = \omega_{pe}^2 = \text{const} > 0$; $\omega_{pe}$ is the Langmuir frequency for $\rho = 1$; $\beta > 0$ is the interface velocity. The profile function $\rho(\xi) \geq 0$ is 1 at $\xi \to -\infty$ and vanishes at $\xi \to +\infty$. We assume that the vector potential magnitude is small, $|A(t,x)| \ll 1$, so that we neglect a longitudinal electric current generation, including the recoil effects and magnetic field induction. For simplicity, the spatial coordinate is normalized to the speed of light in vacuum ($x \to x/c$) so that everywhere below $c = 1$.

We are interested in the solutions of Eq. (12) which have a form of a combination of incident waves coming from $x \to +\infty$, transmitted waves going to $x \to -\infty$, and reflected waves propagating to $x \to +\infty$.

In order to analyse the parameters of the reflected and transmitted waves, in particular, to find the reflection and transmitted coefficients, we use an idealized infinitely sharp step-like profile

$$(13) \quad \rho_\theta(x - \beta t) = \theta(\beta t - x),$$

where $\theta$ is the Heaviside step function, $\theta(\xi) = 0$ for $\xi < 0$ and $\theta(\xi) = 1$ for $\xi > 0$. In this case Eq. (12) for $x > \beta t$ has the form

$$(14) \quad A_{tt} - A_{xx} = 0,$$

its general solution is $A(t,x) = f(x - t) + g(x + t)$, where $f$ and $g$ are arbitrary differentiable functions. For $x < \beta t$, Eq. (12) becomes

$$(15) \quad A_{tt} - A_{xx} + n\,A = 0,$$

its bounded general solution can be cast in the form

$$(16) \quad A(t,x) = \int_{-\infty}^{+\infty} a(\kappa) e^{-i\left(\sqrt{\kappa^2+n}\,t - \kappa x\right)} d\kappa,$$

where $a(\kappa)$ is an arbitrary integrable function.

If the incident wave is a plane wave $g(x + t) = \exp(-i\omega(t + x))$, the reflected and transmitted waves are also plane waves, since Eq. (12) is linear. Thus, for the reflected wave: $f(x - t) = a_R \exp(-i\omega_R(t - x))$. On the left from the interface, the transmitted wave corresponds to $a(\kappa) = a_T \delta(\kappa - k_T)$ with $k_T$ the wavenumber and the Dirac delta function $\delta$. Consequently, we write the solution as a piecewise function

$$(17) \quad A(t,x) = \begin{cases} \exp(-i\omega(t+x)) + a_R \exp(-i\omega_R(t-x)), & x > \beta t; \\ a_T \exp(-i(\omega_T t + k_T x)) + a_C \exp(i(\omega_C t - k_C x)), & x < \beta t. \end{cases}$$

The wave parameters $k_T, \omega_T, k_C, \omega_C$ obey the following conditions

$$(18) \quad \omega_T = \sqrt{k_T^2 + n},$$
$$(19) \quad \omega_C = \sqrt{k_C^2 + n}.$$

Here we introduce a co-reflected wave $a_C \exp(i(\omega_C t - k_C x))$ in medium, which appears if the interface is luminal or superluminal. For $0 < \beta < 1$ one can change to a reference frame co-moving with the interface. There the assumption on the presence of the co-reflected wave is definitely superfluous, hence $a_C = 0$. For $\beta \geq 1$, the reflected wave is senseless, because the super-luminal interface cannot induce waves propagating faster than itself. Thus, $a_R = 0$, and may appear the co-reflected wave propagating in the direction of the interface.

The constant parameters $a_R, \omega_R, a_T, k_T, a_C, k_C$ in Eq. (17) are rather easily determined from the condition of a continuity of the solution represented by Eq. (17) and its first-order derivative. In particular, from this condition we obtain

$$(20) \quad a_T = 1 + a_R - a_C,$$
$$(21) \quad \omega_R = \frac{1+\beta}{1-\beta}\omega.$$

### 3.1. Subluminal interface, $0 < \beta < 1$

For the subluminal interface, $0 < \beta < 1$, the co-reflected wave is superfluous: $a_C = 0$. In order to induce non-evanescent transmitted wave, the incident wave frequency should be sufficiently high:

$$(22) \quad \omega^2 > \omega_{cr}^2,$$
$$(23) \quad \omega_{cr} = \sqrt{n(1-\beta)/(1+\beta)}.$$

The condition that $k_T$ and $\omega_T$ have same sign, $k_T \omega_T > 0$, leads to a higher threshold:

$$(24) \quad \omega^2 > \omega_{st}^2 > \omega_{cr}^2,$$
$$(25) \quad \omega_{st} = \sqrt{n}/(1+\beta).$$

The reflected wave magnitude is:

$$(26) \quad a_R = \frac{\omega - \sqrt{\omega^2 - n(1-\beta)/(1+\beta)}}{\omega + \sqrt{\omega^2 - n(1-\beta)/(1+\beta)}}.$$

The parameters of the transmitted wave are:

$$(27) \quad k_T = \frac{\sqrt{\omega^2 - n(1-\beta)/(1+\beta)} - \beta\omega}{1-\beta}, \quad \omega_T = \frac{\omega - \beta\sqrt{\omega^2 - n(1-\beta)/(1+\beta)}}{1-\beta}, \quad a_T = \frac{2\omega}{\omega + \sqrt{\omega^2 - n(1-\beta)/(1+\beta)}}.$$

For large $\omega$, quite expectedly, the reflection becomes negligible and the incident wave simply becomes the transmitted wave at the interface; thus, the transmitted wave phase velocity is negative (i.e. the phase propagates in space from $+\infty$ to $-\infty$). When $\omega$ decreases to the greatest threshold of the inequality (24), the transmitted wave degenerates to a non-travelling standing wave:

$$(28) \quad \omega = \omega_{st}: k_T = 0, \quad \omega_T = \sqrt{n}, \quad a_T = 2/(1+\beta),$$
$$\omega_R = \sqrt{n}/(1-\beta), \quad a_R = (1-\beta)/(1+\beta).$$

In the interval of $\omega_{st}^2 > \omega^2 > \omega_{cr}^2$, the wavenumber $k_T$ becomes negative, $k_T < 0$; thus, the transmitted wave phase velocity becomes positive (i.e. the phase propagates in space from $-\infty$ to $+\infty$). This causes a terminology problem: the formally transmitted wave propagating in the direction opposite to the incident radiation can be regarded as an *internally reflected* wave.

When $\omega$ riches the threshold of Eq. (22), plasma becomes opaque:

$$(29) \quad \omega = \omega_{cr}: k_T = -\beta\sqrt{n/(1-\beta^2)}, \quad \omega_T = \sqrt{n/(1-\beta^2)}, \quad a_T = 2,$$
$$\omega_R = \sqrt{n(1+\beta)/(1-\beta)} = n/\omega, \quad a_R = 1.$$

For $\omega$ in the interval $\omega^2 < \omega_{cr}^2$, the wavenumber $k_T$ and frequency $\omega_T$ are complex numbers, i.e., the transmitted wave is evanescent while the reflection coefficient $\mathcal{R}_{\text{sub}} = |a_R|^2$ remains to be 1.

We emphasize that the appearance of the standing wave and the transmitted wave with positive phase velocity is impossible if $\beta = 0$ (unmovable interface), since in that case both the thresholds in Eq. (24) are equal to each other. When $\omega_{st}^2 > \omega^2 > \omega_{cr}^2$, both the reflected and transmitted waves propagate in the same direction.

In the limit of $\beta \to 1 - 0$,

$$(30) \quad \omega_R = \frac{2\omega}{1-\beta} + O(1), \quad a_R = \frac{n}{8\omega^2}(1-\beta) + O(1-\beta)^2,$$

$$(31) \quad k_T = \omega - \frac{n}{4\omega} + O(1-\beta), \quad \omega_T = \omega + \frac{n}{4\omega} + O(1-\beta), \quad a_T = 1 + \frac{n}{8\omega^2}(1-\beta) + O(1-\beta)^2.$$

We see that while the frequency of the reflected wave increases, its magnitude vanishes, so that

$$(32) \quad a_R \omega_R \to n/(4\omega).$$

The reflection coefficient at the interface is

$$(33) \quad \mathcal{R}_{\text{sub}} = |a_R|^2 = \left|\frac{\omega - \sqrt{\omega^2 - n(1-\beta)/(1+\beta)}}{\omega + \sqrt{\omega^2 - n(1-\beta)/(1+\beta)}}\right|^2.$$

When the interface velocity tends to the speed of light in vacuum, the reflection coefficient vanishes as

$$(34) \quad \mathcal{R}_{\text{sub}} \approx \frac{n^2(1-\beta)^2}{64\omega^4}, \quad \beta \to 1 - 0,$$

which is the same limit as Eq. (5) if $\omega_i$ is replaced by $\omega$.

The solution for an arbitrary incident waveform $A_{in}(x+t)$ can be constructed as follows. First, we represent the incident waveform as a sum of plane waves,

$$(35) \quad A_{in}(x+t) = \int_{-\infty}^{+\infty} a_I(\omega) e^{-i\omega(t+x)} d\omega,$$

This is possible as long as the Fourier transform of $A_{in}$ exists. Then we define $a_R$, $\omega_R$, $a_T$, $\omega_T$, $k_T$ according to Eqs. (18), (21), (26), (27) as functions of $\omega$. Finally, we obtain

$$(36) \quad A(t,x) = \int_{-\infty}^{+\infty} a_I(\omega) \left[e^{-i\omega(t+x)} + a_R(\omega) e^{-i\omega_R(t-x)}\right] d\omega \quad \text{for} \quad x > \beta t;$$

$$(37) \quad A(t,x) = \int_{-\infty}^{+\infty} a_I(\omega) a_T(\omega) e^{-i(\omega_T t + k_T x)} d\omega \quad \text{for} \quad x < \beta t.$$

Obviously, the functions defined by Eqs. (36) and (37) obey Eqs. (14) and (15), respectively, while the choice of $a_R$, $\omega_R$, $a_T$, $\omega_T$, $k_T$ ensures the continuity of the overall solution and its first-order derivatives at the interface.

### 3.1.1. Numerical solution

For illustrating the electromagnetic wave interaction with the moving plasma-vacuum interface we numerically solve Eq. (12). We discretize Eq. (12) according to the finite-difference time-domain (FDTD) method (e.g. see Ref. [15]). Then the finite-difference equation is solved numerically with an initial condition representing the incident electromagnetic wave packet and with the periodic boundary condition.

The spatial coordinate is normalized to the speed of light in vacuum $x \to x/c$, so that its dimension transforms into that of time, $[x/c] = $ second. As a result, all the 3 additive terms in Eq. (12) have the same dimension, namely that of the product of the vector-potential and the time in the negative second power, $[At^{-2}] = $ Volt $\times$ meter$^{-1}$ $\times$

second$^{-1}$. Thus, there is no restriction on the independent normalization of the vector-potential and time. In other words, the reader can choose the desired units for the vector-potential and for time independently, respecting the approximation assumptions following Eq. (12).

The time step $dt$ is constrained by the Courant–Friedrichs–Lewy (CFL) condition for the numerical stability, $dt < dx$, where $dx$ is the spatial mesh size. We choose $dt = 0.999 dx$ and present simulations for $dx = 1/128$ and $dx = 1/64$ where appropriate based on the necessity of resolving high frequencies.

In the simulations, the parameter $n=10$ (it characterizes the plasma electron density). We use two versions of the profile function $\rho(\cdot)$: one is a plasma-vacuum interface,

$$(38) \quad \rho_\infty(x,t) = \frac{1}{2}\left[1 - \tanh\left(\frac{x - X_M - \beta t}{l_{\text{slope}}}\right)\right],$$

and another is a plasma slab with the thickness of $l = 20$,

$$(39) \quad \rho_{20}(x,t) = \frac{1}{2}\left[\tanh\left(\frac{x + l - X_M - \beta t}{l_{\text{slope}}}\right) - \tanh\left(\frac{x - X_M - \beta t}{l_{\text{slope}}}\right)\right].$$

Here $X_M = 0$ is the initial position of the leading interface of the profile function; $l_{\text{slope}} = 1/5$ is the width of the front and rear transition regions (slopes) of the slab; $\beta$ is the velocity of the interface or slab.

We use two versions of the electromagnetic wave packet: one is a relatively long wave packet with the duration of $\tau = 50$, an exponential tail width of $\tau_{\text{tail}} = 5$, and the characteristic frequency of $\omega$,

$$(40) \quad A_5(x,t) = \frac{1}{2}\left[\tanh\left(\frac{x + t - X_P}{\tau_{\text{tail}}}\right) - \tanh\left(\frac{x + t - \tau - X_P}{\tau_{\text{tail}}}\right)\right]\sin[\omega(x+t)].$$

Another version is a short pulse with the duration (length) $L = 1$,

$$(41) \quad A_1(x,t) = \exp\left[-\frac{1}{2}\left(\frac{x + t - X_P}{L}\right)^2\right].$$

Here $X_P = 50$ is the initial position of the front of the long pulse or the center of the short pulse. We are interested in the electric field. Under the approximation assumptions following Eq. (12), we neglect the longitudinal electron motion in plasma, so the electric field potential is zero, therefore,

$$(42) \quad E = -\partial A/\partial t.$$

In the case of the short pulse the electric field is

$$(43) \quad E_1(x,t) = \frac{x + t - X_P}{L^2}\exp\left[-\frac{1}{2}\left(\frac{x + t - X_P}{L}\right)^2\right].$$

Its spatial Fourier transform exhibit a continuous localized spectrum with the characteristic wavenumber of $k_1 \sim 1/L$,

$$(44) \quad \widetilde{E_1}(k,t) = ikL\exp\left[-\frac{1}{2}L^2 k^2 - ik(t - X_P)\right].$$

The point of collision of the pulse front or center with the plasma interface is $X_* = (X_M + \beta X_P)/(1 + \beta)$.

In this particular subsection, the plasma-vacuum interface motion is subluminal, $\beta = 0.75$. This gives the critical frequency at the threshold of plasma transparency/opaqueness according to Eq. (22): $\omega_{cr} \approx 1.195$. The frequency which induces a standing wave is $\omega_{st} \approx 1.807$, according to Eq. (24). The frequency upshift factor is 7, according to Eqs. (2), (21). The collision point is at $X_* \approx 21.43$.

Fig. 2 shows the interaction of the long monochromatic electromagnetic pulse, Eq. (40), with the subluminal plasma-vacuum interface, given by Eq. (38), for the three cases: when the incident radiation is of critical frequency, $\omega_{cr}$, when it is of frequency $\omega_{st}$, and when it is $\omega = 3$, in terms of the electromagnetic field strength. In the first case, Fig. 2 (a), the incident radiation gets almost reflected while some portion is transmitted with positive phase and group velocity; it leaks out of plasma behaving as a reflected radiation. In the second case, Fig. 2 (b), the reflection is much weaker while transmitted radiation becomes a slowly spreading standing wave with zero group velocity and infinite phase velocity. Sooner or later this standing wave should break the approximation that allows neglecting longitudinal motion of electrons. This can be a mechanism of formation of Langmuir waves [16,17] and electromagnetic solitons [18]. In the third case, Fig. 2 (c), the reflection is even weaker, while the transmitted radiation looks like a classical refracted ray.

Fig. 3 shows how the short electromagnetic pulse, Eq. (43), interacts with the moving plasma-vacuum interface, Eq. (38), in terms of the electromagnetic field strength distribution in the plane $(x, t)$. The short pulse has a relatively wide spectrum, therefore its frequencies in different spectral regions produce different effects corresponding to monochromatic incident radiation with certain frequency. Relatively low frequencies are efficiently reflected with the double Doppler upshift. Sufficiently high frequencies are transmitted; the frequency $\omega_{st}$ induces a standing wave; frequencies higher or lower than this value produce, respectively, the transmitted waves with negative or positive phase and group velocity.

Further details on the short pulse interaction are seen in Fig. 4, presenting the profiles of the electron density and the electromagnetic field strength. For the greatest time shown we draw the local wavenumber of the electromagnetic radiation. It is computed as an inverse distance between adjacent local extrema of the electric field (therefore it always shows the maximum frequency stored in the field). We note the dispersion effects in the transmitted radiation (for the comprehensive description of the plasma dispersion effects on the electromagnetic wave evolution in plasmas see Ref. [19]). Due to the dispersion effects the transmitted radiation acquires the form of a long train of waves with varying wavenumber and amplitude. At the collision point $X_*$ the wavenumber of the transmitted wave vanishes, as marked by the vertical dashed arrow.

The reflected radiation takes the form of a relatively long almost monochromatic wave packet with the leading short peak and a long weak tail, Fig. 5. The wavenumber corresponds to the upshifted frequency $\omega_R \approx 8.4$, which implies the incident frequency of $\omega \approx 1.2$. We note that the portion of the transmitted radiation with positive phase velocity (which is greater than the speed of light in vacuum) exists at the interface for a long time after the pulse collision with the interface, Fig. 4. It actually behaves as reflected radiation. Leaving plasma, it forms the long weak tail of the reflected wave packet seen in Fig. 5. It is the most striking consequence of the dispersion effects and moving interface: the appearance of two types of reflected radiation, one at the interface and another inside plasma, in the form of a long wavepacket spanning from the collision point till the moving interface.

In the case of a finite plasma slab, Eq. (39), new features appear, Figs. 6-8. Fig. 6 presents the electric field strength distribution in the plane $(x, t)$. From $t = 0$ till $t \approx 43$ the electric field distribution in Fig. 6 is exactly the same as in Fig. 3; later they differ due to the refraction of the electromagnetic waves at the rear side of the plasma slab.

The profiles of the electric field strength and the plasma density are shown in Fig. 7. The radiation refraction occurs on both sides of the slab. In particular, we see the second reflection inside the slab: it reveals itself at later time, together with the emission into vacuum of the transmitted waves (stored in the slab) with positive phase velocity, Fig. 8. The second reflection leaked from the plasma slab has a higher frequency than the first, but its magnitude is quite small. We note that due to multiple reflection, the maximum local wavenumber is not informative enough, therefore we show also the local spectral density. It is computed in every point $x_0$ as the fast Fourier transform of the electric field $E(x)$ multiplied by the Gaussian centered at that point with the characteristic width of $\sigma = 10$: $I_k(x_0) = \left| \text{FFT}_x \left[ E(x) e^{-((x-x_0)/\sigma)^2} \right](k) \right|$. The appearance of the second reflection is seen in Fig. 7 (d) as the local spectral maximum at $k \approx 7$ for $x = 60$ (inside the plasma slab).

### 3.2. Superluminal interface, $\beta > 1$

For the superluminal interface, $\beta > 1$, the reflected wave is senseless, $a_R = 0$. The transmitted and the co-reflected waves are non-evanescent for any real $\omega$, because the right-hand part of Eq. (22) is negative. However, the requirement that the signs of $k_T, \omega_T$ and $k_C, \omega_C$ are pairwise the same leads to the same condition on $\omega$ as in Eq. (24).

The parameters of the transmitted wave are:

(45) $\quad a_T = \frac{1}{2} + \frac{\omega}{2\sqrt{\omega^2 + n(\beta-1)/(\beta+1)}}$, $\quad \omega_T = \frac{\beta\sqrt{\omega^2 + n(\beta-1)/(\beta+1)} - \omega}{\beta - 1}$, $\quad k_T = \frac{\beta\omega - \sqrt{\omega^2 + n(\beta-1)/(\beta+1)}}{\beta - 1}$.

The parameters of the co-reflected wave are:

(46) $\quad a_C = \frac{1}{2} - \frac{\omega}{2\sqrt{\omega^2 + n(\beta-1)/(\beta+1)}}$, $\quad \omega_C = \frac{\beta\sqrt{\omega^2 + n(\beta-1)/(\beta+1)} + \omega}{\beta - 1}$, $\quad k_C = \frac{\sqrt{\omega^2 + n(\beta-1)/(\beta+1)} + \beta\omega}{\beta - 1}$.

When $\omega$ riches the greatest threshold of the inequality (24), the transmitted wave degenerates to non-travelling standing wave:

(47) $\quad \omega = \omega_{st}: k_T = 0, \quad \omega_T = \sqrt{n}, \quad a_T = (\beta + 1)/2\beta,$
$k_C = 2\beta\sqrt{n}/(\beta^2 - 1), \quad \omega_C = \sqrt{n}(\beta^2 + 1)/(\beta^2 - 1), \quad a_C = (\beta - 1)/2\beta.$

For $\omega$ in the interval $\omega^2 < \omega_{st}^2$, the wavenumber $k_T$ is negative, $k_T < 0$. Similarly to the case of subluminal interface, the transmitted wave phase velocity is positive, so the transmitted wave propagates in the same direction as the co-reflected wave. Consequently, it can be regarded as an *internally reflected* wave.

When $\beta \to 1 + 0$,

(48) $\quad a_T = 1 - \frac{n}{8\omega^2}(\beta - 1) + O(\beta - 1)^2$, $\quad \omega_T = \omega + \frac{n}{4\omega} + O(\beta - 1)$, $\quad k_T = \omega - \frac{n}{4\omega} + O(\beta - 1)$,

(49) $\quad a_C = \frac{n}{8\omega^2}(\beta - 1) + O(\beta - 1)^2$, $\quad \omega_C = \frac{2\omega}{\beta - 1} + O(1)$, $\quad k_C = \frac{2\omega}{\beta - 1} + O(1)$.

As in the case of the subluminal interface, the co-reflected wave frequency increases while its magnitude vanishes, so that

(50) $\quad a_C \omega_C \to n/(4\omega)$.

We can treat the value $|a_C|^2$ as a co-reflection coefficient

(51) $\quad \mathcal{R}_{\text{super}} = |a_C|^2 = \left| \frac{1}{2} - \frac{\omega}{2\sqrt{\omega^2 + n(\beta-1)/(\beta+1)}} \right|^2$.

When the interface velocity tends to the speed of light in vacuum, the co-reflection coefficient vanishes as

$$(52) \quad \mathcal{R}_{\text{super}} \approx \frac{n^2(\beta-1)^2}{64\omega^4}, \quad \beta \to 1 + 0.$$

This asymptotic is symmetrical to Eq. (34) with respect to the point $\beta = 1$, and also is similar to the limit in Eq. (5) if $\omega_i$ is replaced by $\omega$.

### 3.2.1. Numerical solution

In this subsection, the plasma-vacuum interface motion is superluminal, $\beta = 1.25$. The plasma is transparent for any incident frequency, according to Eq. (22). The frequency which induces a standing wave is $\omega_{st} \approx 1.405$, according to Eq. (24). The frequency upshift factor is 9, according to Eqs. (2), (21). The collision point is at $X_* \approx 27.78$.

Fig. 9 shows the interaction of the long monochromatic electromagnetic pulse, Eq. (40), with the superluminal plasma-vacuum interface, Eq. (38), for the three cases of the incident radiation frequency: $\omega = 0.5$, $\omega = \omega_{st}$, and $\omega = 3$. As expected, the radiation induced by the incident pulse remains behind the superluminal interface. We see the co-reflected wave and the transmitted wave. In the case of $\omega < \omega_{st}$, the co-reflected and transmitted waves propagate in the same direction; it looks like there are two reflected waves. When $\omega = \omega_{st}$, the transmitted wave becomes stationary. For $\omega > \omega_{st}$, the co-reflected and transmitted wave move in opposite directions.

Fig. 10 shows the case for the short electromagnetic pulse given by Eq. (43). In contrast to the case of the long monochromatic pulse, here the transmitted wave is not detached from the co-reflected wave. The wide spectrum of the incident short pulse contains a frequency for each wavenumber between the maximum wavenumber of the co-reflected wave and the minimum wavenumber of the transmitted wave.

The profiles of plasma density and electric field strength as well as the local wavenumber are shown in Fig. 11. The incident short pulse induces in plasma a long wavepacket with varying frequency. The local wavenumber looks symmetrical with respect to the collision point, $x = X_* \approx 27.78$, except the front of the wave packet at $90 < x < 100$, where it sharply grows. This region can be regarded as the location of the co-reflected pulse. Later in time the induced wavepacket becomes longer, its magnitude decreases, the part corresponding to the co-reflected pulse (where the local wavenumber sharply grows) elongates, yet the maximum wavenumber remains the same. This is seen in Fig. 12, in terms of the dependence of $\zeta = x - t$, so that $\zeta = 0$ corresponds to the front of the wavepacket. The local wavenumber is well approximated by the exponential integral E function of the order 1: $E_1(-\zeta/290)$, see Fig. 12(b). The product of the local wavenumber and the electric field envelope magnitude remains finite, Fig. 12(a), and slowly decreases with time, with the same rate as the electric field magnitude.

The case of the plasma slab, Eq. (39), is shown in Fig. 13-16. The electric field strength distribution in the plane $(x, t)$ presented in Fig. 13 is exactly the same as in Fig. 10 in the region below the world line of the rear side of the superluminal slab. Unlike the case of the infinite plasma layer in Fig. 10, here we see the separation of the co-reflected pulse and transmitted radiation. The transmitted pulse in vacuum region behind the slab is a wavepacket with the gradually decreasing frequency, Fig. 14. The frequency falls almost to zero, which leads to an appearance of a relatively long interval of an almost constant non-zero vector potential $A$. This is seen in Fig. 15(c) which represents the same interaction as Fig. 14 but in terms of the vector potential. Fig. 16 shows that the co-reflected pulse in terms of the electric field strength consists of two antisymmetric sub-pulses; the thicker the slab the longer the co-reflected pulse, while the maximum frequency is the same. The leading sub-pulse has negative chirp, while the trailing sub-pulse has positive chirp. The former is created at the front side and the latter – at the rear side of the slab.

### 3.3. Luminal interface, $\beta = 1$

When the medium-vacuum interface moves with the speed of light in vacuum, $\beta = 1$, the limits in the previous subsections suggest the following parameters of Eq. (17):

$$(53) \quad a_T = 1, \quad \omega_T = \omega + \frac{n}{4\omega}, \quad k_T = \omega - \frac{n}{4\omega}, \quad a_R = 0, \quad a_C = 0.$$

However, the left and right values of the first-order derivatives of $A(t, x)$ defined by Eq. (17) at the interface do not match:

$$(54) \quad A_t|_{x<\beta t} = -i\left(\omega + \frac{n}{4\omega}\right), \quad A_t|_{x>\beta t} = -i\omega.$$

$$(55) \quad A_x|_{x<\beta t} = -i\left(\omega - \frac{n}{4\omega}\right), \quad A_x|_{x>\beta t} = -i\omega.$$

In other words, the wave phase derivatives cannot be continuous in this case.

We return to Eq. (12), now for $\beta = 1$:

$$(56) \quad A_{tt} - A_{xx} + n\,\rho(x - t)A = 0$$

with the same assumption about the profile $\rho$: $\rho \geq 0$, $\rho(-\infty) = 1$, and $\rho(+\infty) = 0$. Changing variables (see Fig. 17) in this equation to

$$(57) \quad \xi = \frac{1}{2}n\,S(x - t), \quad \eta = \frac{1}{2}(x + t),$$

where $S$ is the antiderivative of $\rho$,
$$(58) \quad S'(z) = \rho(z),$$
we obtain
$$(59) \quad A_{\xi\eta} = A.$$
The function $S$ is monotonic because its derivative $\rho$ is non-negative. The Fourier transform with respect to $\eta$,
$$(60) \quad a(\xi,\Omega) = \frac{1}{\sqrt{2\pi}} \int_{-\infty}^{+\infty} A(\xi,\eta) e^{i\Omega\eta} d\eta,$$
reduces the partial differential equation (59) with respect to two variables $\xi$ and $\eta$ to the ordinary differential equation with respect to a single variable $\xi$:
$$(61) \quad a'(\xi) = \frac{i}{\Omega} a(\xi),$$
where we treat $\Omega$ as a parameter in the definition of $a$ in Eq. (60). The solution is
$$(62) \quad a(\xi,\Omega) = a_0(\Omega) \exp[i\,\xi/\Omega].$$
Here $a_0(\Omega)$ is determined by the initial condition. Finally, taking inverse Fourier transform of Eq. (62) with respect to $\Omega$, we obtain the general solution of Eq. (59):
$$(63) \quad A(\xi,\eta) = \frac{1}{\sqrt{2\pi}} \int_{-\infty}^{+\infty} a_0(\Omega) \exp\left[-i\left(\Omega\eta - \frac{\xi}{\Omega}\right)\right] d\Omega.$$
This integral certainly converges if the function $a_0(\Omega)$ is bounded and vanishes at infinity sufficiently fast. The expression can be cast in the form
$$(64) \quad A(\xi,\eta) = \frac{1}{\sqrt{2\pi}} \sqrt{-\frac{\xi}{\eta}} \int_{-\infty}^{+\infty} a_0\left(\sqrt{-\frac{\xi}{\eta}} u\right) \exp\left[-i\sqrt{-\xi\eta}\left(u + \frac{1}{u}\right)\right] du.$$

### 3.3.1. Sharp interface and plane wave
If the incident wave is a plane wave, $A_{0w}(\xi,\eta) = \exp(-2i\omega\eta) = \exp(-i\omega(x+t))$, its frequency-domain representation is $a_{0w}(\Omega) = \sqrt{2\pi}\delta(\Omega - 2\omega)$, therefore the solution becomes
$$(65) \quad A_w(t,x) = \exp\left[-i\left(\omega(x+t) + \frac{n}{4\omega}\int_{x-t}^{+\infty}\rho(\xi')d\xi'\right)\right].$$
This solution has the following asymptotics:
$$(66) \quad A_w = \exp[-i\omega(x+t)], \quad x \to +\infty;$$
$$(67) \quad A_w = \exp\left[-i\left(\left(\omega + \frac{n}{4\omega}\right)t + \left(\omega - \frac{n}{4\omega}\right)x + \phi_0\right)\right], \quad x \to -\infty.$$
The latter asymptotic is valid if exists a finite phase shift $\phi_{0w}$:
$$(68) \quad \phi_{0w} = \frac{n}{4\omega}\left(\int_{-\infty}^{+\infty}\{\rho(\xi) - \theta(-\xi)\}d\xi\right).$$
The asymptotic Eq. (67) for zero $\phi_{0w}$ coincides with the limits of Eqs. (31) and (48) for $a_T$, $k_T$ and $\omega_T$. In particular, in the case of $\rho(\xi) = \theta(-\xi)$, we have $\phi_{0w} = 0$ and obtain the discontinuities in the first-order derivatives described by Eqs. (54) and (55). We see that there is only the transmitted wave with parameters described by Eq. (53). The transmitted wave degenerates to a standing wave for
$$(69) \quad \omega = \sqrt{n}/2: \ k_T = 0, \ \omega_T = \sqrt{n}, \ a_T = a_0.$$
For $\omega^2 > n/4$, the transmitted phase velocity is negative (the phase propagates in space from $+\infty$ to $-\infty$). For $\omega^2 < n/4$, the transmitted phase velocity is positive (the phase propagates in space from $-\infty$ to $+\infty$); in this case the transmitted wave can be regarded as *internally reflected* wave. Its frequency, $\omega_{R(T)}$, follows from Eq. (53),
$$(70) \quad \omega_{R(T)} = \left(1 + \frac{n}{4\omega^2}\right)\omega = \left(1 + \frac{n}{4n_{cr}}\right)\omega > 2\omega.$$
Surprisingly, the lower the incident frequency, the greater the frequency of the internally reflected wave.

### 3.3.2. Limit existence
The presence of discontinuities in the solution for the case of sharp interface may cast doubts on the existence of the limit $\beta \to 1$. We again return to Eq. (12),
$$(71) \quad A_{tt} - A_{xx} + n\,\rho(x - \beta t)A = 0,$$
and use the following variables
$$(72) \quad \bar{\xi} = \frac{1}{2}(x - \beta t), \ \eta = \frac{1}{2}(x + t),$$
which transforms the Eq. (71) into
$$(73) \quad (1-\beta^2)A_{\bar{\xi}\bar{\xi}} + 2(1+\beta)A_{\bar{\xi}\eta} - 4n\,\rho(\bar{\xi})A = 0.$$
Further, applying the Fourier transform (with respect to $\eta$) to this equation and using the definition of Eq. (60) we obtain the ordinary differential equation
$$(74) \quad (1-\beta^2)a''(\bar{\xi}) - 2i\Omega(1+\beta)a'(\bar{\xi}) - 4n\rho(\bar{\xi})a(\bar{\xi}) = 0.$$
In the limit of $\beta \to 1$, this is a singular differential equation with a small parameter $(1-\beta^2)$ at the high-order derivative with respect to the variable $\bar{\xi}$. According to the Tikhonov theorem [20], in the limit of $(1-\beta^2) \to 0$ the solution of Eq. (73) tends to the solution of the degenerate equation
$$(75) \quad -i\Omega a'(\bar{\xi}) - n\,\rho(\bar{\xi})a(\bar{\xi}) = 0,$$
where the variable $\bar{\xi}$ becomes $\bar{\xi} = \frac{1}{2}(x - t)$. Changing to $\xi = n\,S(\bar{\xi})$, where $S$ is the antiderivative of $\rho$, we obtain the same equation as Eq. (61). Thus, the general solution given by Eq. (63) is the limit of the general solution of Eq.

(74) for $\beta \to 1$.

### 3.3.3. Eikonal of transmitted plane wave

The plane wave solution Eq. (65) is characterized by its phase (also called eikonal)

$$(76) \quad \Psi_w(t,x) = \omega(x+t) + \frac{n}{4\omega}\int_{x-t}^{+\infty} \rho(\xi')d\xi'.$$

The corresponding wave frequency $\Omega_w$ and wave number $K_w$ are formally defined as

$$(77) \quad \Omega_w = -\partial\Psi_w(t,x)/\partial t = -\omega - \frac{n}{4\omega}\rho(x-t),$$

$$(78) \quad K_w = \partial\Psi_w(t,x)/\partial x = \omega - \frac{n}{4\omega}\rho(x-t).$$

For $x \to +\infty$, these entities are equal and represent an electromagnetic wave in vacuum, whose phase and group velocities are both equal to the speed of light in vacuum. When $\rho(x-t) = \rho > 0$, we have

$$(79) \quad \Omega_w = -\omega - \frac{n\rho}{4\omega}, \quad K_w = \omega - \frac{n\rho}{4\omega}, \quad \Omega_w^2 - K_w^2 = n\rho.$$

In this case the phase and group velocities of the wave are

$$(80) \quad \beta_{ph,w} = \frac{\Omega_w}{K_w} = \frac{n\rho+4\omega^2}{n\rho-4\omega^2}, \quad \beta_{g,w} = \frac{\partial\Omega_w}{\partial K_w} = \frac{n\rho-4\omega^2}{n\rho+4\omega^2}.$$

The former is greater and the latter is less than the speed of light in vacuum (in terms of magnitude), while

$$(81) \quad \beta_{ph,w}\beta_{g,w} = 1.$$

Both velocities have the same sign. The internally reflected radiation corresponds to the case of $\omega^2 < n\rho/4$ that is the condition of positive phase and group velocity.

### 3.3.4. Asymptotic transmission of a short pulse

We consider a short pulse propagation through the luminal plasma-vacuum interface. The short pulse propagates from vacuum at $x \to +\infty$ to plasma at $x \to -\infty$. It is modelled by the Gaussian profile (see Eq. (41))

$$(82) \quad A_p(\eta) = e^{-2\eta^2/L^2} = \exp\left[-\frac{(x+t)^2}{2L^2}\right],$$

where $L$ is the characteristic duration of the pulse. The Fourier transform of this profile according to Eq. (60) gives

$$(83) \quad a_p(\Omega) = \frac{L}{2}e^{-L^2\Omega^2/8},$$

revealing the characteristic frequency $\Omega_p \sim 2/L$.

The long-time asymptotic behaviour of the transmitted wave can be revealed by using the stationary phase method in estimation of the integral of Eq. (63), which takes the form

$$(84) \quad A_{pT}(\xi,\eta) = \frac{1}{\sqrt{2\pi}}\int_{-\infty}^{+\infty} a_p(\Omega)\cos(\Psi)\,d\Omega, \quad \Psi = \xi/\Omega - \Omega\eta.$$

Here we utilized the fact that the function $a_p(\Omega)$ is even, thus the imaginary part vanishes.

The integral can be approximated as

$$(85) \quad A_{pT}(\xi,\eta) \approx \frac{1}{\sqrt{2\pi}}a_p(\Omega_*)\int_{-\infty}^{+\infty}\cos\left[\Psi(\Omega_*) + \frac{1}{2}\Psi''(\Omega_*)(\Omega - \Omega_*)^2\right]d\Omega,$$

where $\Omega_*$ makes the phase stationary:

$$(86) \quad \Psi'(\Omega_*) = -\xi/\Omega_*^2 - \eta = 0 \Rightarrow \Omega_* = \sqrt{-\xi/\eta}.$$

Under the integral in Eq. (84), the quadratic term of the phase leads to rapid oscillations, thus effectively cancelling the integral for frequencies $\Omega$ far from $\Omega_*$. The approximation necessitates that the value of $|\Psi''(\Omega_*)|$ is sufficiently large and that the function $a_p$ is slowly varying near $\Omega_*$.

We note that $\Omega_*$ is real for $\xi < 0, \eta > 0$, and exactly this quadrant is of physical interest: only here the future of the interaction can occur, Fig. 17.

Integrating Eq. (85), we obtain

$$(87) \quad A_{pT}(\xi,\eta) \approx \frac{L}{4}e^{-(L^2/8)|\xi/\eta|}(-\xi)^{1/4}\eta^{-3/4}\left(\cos\left[2\sqrt{-\xi\eta}\right] - \sin\left[2\sqrt{-\xi\eta}\right]\right),$$

which is valid when

$$(88) \quad |\Psi''(\Omega_*)| = 2\eta^{3/2}(-\xi)^{-1/2} \gg 1.$$

We note that this approximation holds for any profile of plasma-vacuum interface that provides an acceptable change of variables in Eq. (57). In terms of $t$ and $x$, the phase of oscillations and the argument of the exponent are

$$(89) \quad 2\sqrt{-\xi\eta} = \sqrt{-nS(x-t)(x+t)}, \quad \frac{\xi}{\eta} = \frac{nS(x-t)}{x+t}.$$

In the simplest case of a sharp interface, we have: $\rho(\xi) = \theta(-\xi)$ and $S = \xi\,\theta(-\xi)$. Then, the frequency and wavenumber of the transmitted radiation are easily obtained from the phase of the oscillatory term of Eq. (87):

$$(90) \quad \Psi_{pT} = 2\sqrt{-\xi\eta} = \sqrt{n}\sqrt{t^2 - x^2},$$

$$(91) \quad \Omega_{pT} = -\partial\Psi_{pT}(t,x)/\partial t = -\sqrt{n}t/\sqrt{t^2 - x^2},$$

$$(92) \quad K_{pT} = \partial\Psi_{pT}(t,x)/\partial x = -\sqrt{n}x/\sqrt{t^2 - x^2}.$$

The frequency and wave number are related to each other, as expected,
$$(93) \quad \Omega_{pT}^2 - K_{pT}^2 = n.$$
The phase and group velocities of the radiation are
$$(94) \quad \beta_{ph,pT} = \frac{\Omega_{pT}}{K_{pT}} = \frac{t}{x}, \quad \beta_{g,pT} = \frac{\partial \Omega_{pT}}{\partial K_{pT}} = \frac{x}{t},$$
$$(95) \quad \beta_{ph,pT} \beta_{g,pT} = 1.$$

Sufficiently far from the interface, both velocities become negative, i.e. both the phase and envelope propagate in space from $+\infty$ to $-\infty$ (in the same direction as the incident short pulse). This corresponds to a (properly) transmitted portion of the incident pulse.

For $x = 0$, the group velocity is zero and the phase velocity is infinite. This is similar to sub- and superluminal cases, when $k_T = 0$. We note that the case $x = 0$ requires $t \gg 1$ in order to satisfy Eq. (88).

For the limit of
$$(96) \quad \xi = \frac{1}{2}n(x - t) \to -0 \quad \text{and} \quad \eta = \frac{1}{2}(x + t) \to +\infty$$
formula (87) predicts the existence of the radiation near the interface, long after the short pulse collided with it. Here the phase and group velocities are positive, i.e. both the phase and envelope propagate in space from $-\infty$ to $+\infty$ (in the opposite direction to the incident short pulse). This corresponds to an *internally reflected* portion of the incident pulse (thus, in the notations $\Omega_{pT}$ and $A_{pT}$ the subscript $T$ for "transverse" becomes misleading). At the limit given by Eq. (96), from Eq. (91) we obtain (replacing the subscript $T$ with more relevant $R$ for "reflected")
$$(97) \quad |\Omega_{pR}| \approx (n/2)\eta^{1/2}(-\xi)^{-1/2} = \frac{1}{2}n^{1/2}(t-x)^{-1/2}(t+x)^{1/2}.$$
Thus, near the interface the radiation frequency increases unboundedly in time, $\Omega_{pR} \to \infty$. From Eq. (87) it follows that the magnitude of the radiation decreases as
$$(98) \quad |A_{pR}| \sim \frac{L}{4}(-\xi)^{1/4}\eta^{-3/4} = 2^{-3/2}Ln^{1/4}(t-x)^{1/4}(t+x)^{-3/4}.$$
Nevertheless, the product of frequency and magnitude increases:
$$(99) \quad |A_{pR}\Omega_{pR}| \sim (nL/8)(-\xi\eta)^{-1/4} = 2^{-5/2}Ln^{3/4}(t^2 - x^2)^{-1/4}.$$

As concerns the tendency of the reflected/co-reflected wave frequency and magnitude, the situation is somehow analogous to that of the sub- and superluminal interfaces: frequency tends to infinity, magnitude – to zero. However, it drastically differs with respect to the tendency of the product of the wave frequency and magnitude: here this product tends to infinity. Moreover, in previous sections, the tendencies for the reflected/co-reflected wave frequency and magnitude are for $\beta \to 1$, whereas here $\beta = 1$ and the tendencies are for $x \to t$ and $t \to \infty$. In other words, here frequency of radiation near the interface increases with time spontaneously.

We emphasize the difference of short and long pulses for the stationary phase approximation: a very long almost monochromatic pulse has an extremely localized Fourier transform (i.e., the expression under the integral (84) is zero almost everywhere), therefore it may be improbable for the phase to become stationary at the pulse characteristic frequency. On the other hand, in this case the plane wave approximation is more appropriate.

Eq. (59) admits the solution which depends on $\xi$ and $\eta$ as
$$(100) \quad A(\xi, \eta) = f(\mu)g(\nu),$$
$$(101) \quad \mu = -\xi\eta = -\frac{n}{4}S(x-t)(x+t), \quad \nu = -\xi/\eta = -nS(x-t)/(x+t).$$
We note that $S$, the antiderivative of $\rho$, is negative for $x < t$ if $S(+\infty) = 0$, as it follows from its definition. Substituting the ansatz Eq. (100) into Eq. (59), we obtain two ordinary differential equations for $f$ and $g$:
$$(102) \quad \mu^2 f''(\mu) + \mu f'(\mu) + (\mu \pm m^2)f(\mu) = 0,$$
$$(103) \quad \nu^2 g''(\nu) + \nu g'(\nu) \pm m^2 g(z) = 0,$$
where $m$ is a real constant. The first equation is an almost canonical form for Bessel functions, the second – is for power functions. The solution embracing both signs of $\pm m^2$ can be cast in the form of the linear combination of products $f(\mu)g(\nu)$ expressed in terms of $\xi, \eta$:
$$(104) \quad A(\xi, \eta) = C_{\pm m}^{\pm 2m} J_{\pm 2m}(2\sqrt{-\xi\eta})\left[-\frac{\xi}{\eta}\right]^{\pm m} + C_{\pm m}^{\pm 2im} J_{\pm 2im}(2\sqrt{-\xi\eta})\left[-\frac{\xi}{\eta}\right]^{\pm im}.$$
In the special case of $m = 0$, the solution takes the form
$$(105) \quad A(\xi, \eta) = \left[C_0^+ J_0(2\sqrt{-\xi\eta}) + C_0^- Y_0(2\sqrt{-\xi\eta})\right]\left[C_0 + C_* \ln\left(-\frac{\xi}{\eta}\right)\right].$$
Here $J$ and $Y$ is, respectively, the Bessel function of the first and second kind. $C$ denotes an arbitrary constant; its superscript corresponds to the order of the Bessel function; its subscript correspond to the power of the ratio $(-\xi/\eta)$; the signs of the superscript and subscript are independent. In terms of $t$ and $x$ the arguments of the Bessel and power functions are given by Eq. (89).

In the case of an interface-like plasma density profile, i.e. $\rho \geq 0$, $\rho(-\infty) = 1$, and $\rho(+\infty) = 0$, we assume for the antiderivative $S$, $S'(z) = \rho(z)$, the "boundary condition" $S(+\infty) = 0$. Then on the right from the interface the arguments of the Bessel function and positive power function degenerate to zero, while the argument of the negative power function becomes infinite.

When $2m$ is not a negative integer, the solution near zero argument has the approximation
$$(106) \quad J_{2m}\left(2\sqrt{-\xi\eta}\right)\left[-\frac{\xi}{\eta}\right]^{\pm m} \sim \frac{1}{\Gamma(2m+1)}\begin{cases}(-\xi)^{2m} \text{ for } "+m",\\ \eta^{2m} \text{ for } "-m".\end{cases}$$

At infinite argument, the solution has the asymptotic
$$(107) \quad J_{2m}\left(2\sqrt{-\xi\eta}\right)\left[-\frac{\xi}{\eta}\right]^{\pm m} \sim \pi^{-1/2}(-\xi\eta)^{-1/4}\left[-\frac{\xi}{\eta}\right]^{\pm m}\cos(2\sqrt{-\xi\eta}-\pi m-\pi/4),$$
similarly to Eq. (87). For $m=1/2$, the factor before $\cos(\cdot)$ becomes $\pi^{-1/2}(-\xi)^{1/4}\eta^{-3/4}$.

Assuming in Eq. (64) that $a_0(\Omega) = 1$ and using the identity (based on Ref. [21])
$$(108) \quad \text{V.P.}\int_{-\infty}^{+\infty}\cos\left[\alpha\left(u+\frac{1}{u}\right)\right]du = -2\pi J_1(2\alpha),$$
we find that the solution
$$(109) \quad A(\xi,\eta) = -\sqrt{2\pi}\,J_1\left(2\sqrt{-\xi\eta}\right)\sqrt{-\frac{\xi}{\eta}}$$
(for $m=1/2$ in terms of Eq. (104)) is exactly the expression of Eq. (64) for $a_0(\Omega) = 1$, thus it is the solution for the infinitely short incident electromagnetic pulse $A_0(\eta) = \sqrt{2\pi}\,\delta(\eta)$, where $\delta(\cdot)$ is the Dirac delta function. Fig. 18 illustrates the electric field strength $E = -\partial A/\partial t$ corresponding to Eq. (109) in terms of $x$ and $t$. It is singular at $x+t = 0$ and equals $\sqrt{2\pi}$ at $x-t = 0$. We note that Eq. (109) exhibits similar asymptotes in the limit of Eq. (96) as those defined by Eqs. (97)-(99).

### 3.3.5. Spectrum of a transmitted short pulse

From the solution in the form of Eq. (63) we derive the spatial spectrum of the transmitted short pulse in the form of Eq. (82) in terms of the electric field strength, in the simplest case of a sharp interface: $\rho(\xi) = \theta(-\xi)$, $S = \xi\,\theta(-\xi)$. Here, for the sake of brevity, we present a shortened description. First, we express $\xi$ and $\eta$ in Eq. (63) via $x$ and $t$ using Eq. (57), then we multiply the result by $\exp(ikx)$ and perform the Fourier transform with respect to $x$ under the integral; this step concerns only expressions of the form $\exp[-i(\Omega - n/\Omega)x/2]\theta(t-x)$ and $\exp[-i\Omega x/2]\theta(x-t)$; their Fourier transforms are expressed via the sum of the reciprocal function $1/z$ and Dirac delta function $\delta(z)$ with arguments $z = \Omega - k \pm \sqrt{k^2+n}$ or $z = \Omega - 2k$. Then we integrate the outcome with respect to $\Omega$ and obtain the spectrum of the vector-potential $a(k,t) = \frac{1}{\sqrt{2\pi}}\int_{-\infty}^{+\infty}A(x,t)e^{ikx}dx$ as the function of the wavenumber $k$ and time $t$. Finally, we take the time derivative and negate it to obtain the spatial spectrum of the electric field strength:
$$(110) \quad \tilde{E}(k,t) = \sum_\pm \frac{iL}{4}(k\pm K)\exp\left[-\frac{L^2}{8}(k\pm K)^2 \mp itK\right]\text{erfc}\left[\frac{-4t+iL^2(k\pm K)}{2\sqrt{2}\,L}\right] +$$
$$\frac{iLk}{2}\exp\left[-\frac{1}{2}L^2k^2 - itk\right]\left(1+\text{erf}\left[\frac{-2t+iL^2k}{\sqrt{2}\,L}\right]\right),\quad K = \sqrt{k^2+n}.$$

Here the sum symbol means two terms, the first one assumes the upper sign in subexpressions like $\pm$, the second one assumes the bottom sign. For $t \ll -1$, as expected, the spectrum is approximately equal (exponentially close) to
$$(111) \quad \tilde{E}(k, t \ll -1) \approx ikL\exp\left[-\frac{1}{2}L^2k^2 - itk\right],$$
which the same expression as Eq. (44) for $X_p = 0$, corresponding to Eq. (82). For $t \to \infty$, we have
$$(112) \quad \tilde{E}(k, t\to\infty) \approx \frac{nL}{2(k+\sqrt{k^2+n})}\exp\left[it\sqrt{k^2+n}-i\pi/2\right].$$

This approximation is also exponentially close to the exact formula for sufficiently large $k$, where the modulus of the Fourier transform is approximated by $|\tilde{E}(k,t)| \approx nL/(4k)$. Such the spectrum where the dependence on the wavenumber is described by the reciprocal function is characteristic for jumps like the Heaviside step function. We see that the modulus of the right-hand side of Eq. (112) does not depend on time, while the phase depends on wavenumber asymptotically almost linearly. It means that the energy stored in each wavenumber $k$ is constant in time, thus the shape of the corresponding wavepacket in the time domain changes in time only because the phase changes in time. Since the spectrum contains infinitely high wavenumbers, the resulting wavepacket should have weakly oscillating and highly oscillating parts. Fig. 19 shows the modulus of the right-hand side of Eq. (110) at different time and the corresponding wavepackets computed numerically via the inverse Fourier transform of Eq. (110). The wavepacket shape is similar to the exact solution for infinitely short incident pulse seen in Fig. 18 (except the unbounded increase at the left-hand side of the profile in Fig. 18).

### 3.3.6. Numerical solution

In this subsection, the plasma-vacuum interface motion is luminal, $\beta = 1$. The plasma is transparent for any incident frequency, according to Eq. (22). The frequency which induces a standing wave is $\omega_{st} \approx 1.581$, according to Eq. (24). The frequency upshift factor is infinity, according to Eqs. (2), (21). The collision point is at $X_* = 25$.

Fig. 20 shows the interaction of the long monochromatic electromagnetic pulse, Eq. (40), with the luminal plasma-vacuum interface, Eq. (38), for the three cases of the incident radiation frequency: $\omega = 0.5$, $\omega = \omega_{st}$, and $\omega = 3$. It almost resembles Fig. 2 and Fig. 9 except the absence of the reflected radiation. In the case of $\omega = 0.5$, when the transmitted radiation has positive phase and group velocity (thus behaves like a reflected radiation), we see that the radiation frequency is higher as the radiation is closer to the luminal plasma interface. This effect becomes more pronounced with the incident pulse having a wide spectrum, as in the next figure.

Fig. 21 shows how the short electromagnetic pulse given by Eq. (43) interacts with the luminal plasma-vacuum interface. We see that the electric field frequency increases in time near the world line of the interface, while the electric field magnitude does not vanish, in accordance with the prediction in the previous subsection. This behaviour drastically differs from what we see in Fig. 3 and Fig. 10.

The profiles of the electric field strength and electron density as well as the local wavenumber are shown in Figs. 22-24. The electric field strength profiles are expectedly very similar to that in Fig. 19(b,c), which are computed from the exact formula of Eq. (110) for the corresponding spectrum. Similarly to Fig. 11, the transmitted radiation is a long wavepacket negatively chirped till $x = X_*$, then positively chirped. In contrast to the superluminal interface (Figs. 11-12), here the electric field strength magnitude is not zero when the wavenumber is maximum, at the front of the wavepacked, i.e. in the vicinity of the luminal plasma-vacuum interface. Moreover, while the electric field magnitude slowly decreases with time, the corresponding maximum wavenumber increases, Figs. 23-24. The growth near the interface of the wavenumber $k$ and the product of the electric field envelope and the wavenumber $k|E|$ predicted by Eqs. (97), (99) is seen in Fig. 24. We note that according to Eq. (42), the electric field strength is related to the vector potential as $|E| \sim |A\omega|$. Therefore, taking $\zeta = x - t$, if $|A| \sim (-\zeta)^{1/4}$ by Eq. (87) and $|\omega|, k \sim (-\zeta)^{-1/2}$ by Eq. (97), then $|E| \sim (-\zeta)^{-1/4}$, so that $k|E| \sim (-\zeta)^{-3/4}$. The functions behaviour at $x \to t - 0$ is very similar to that of the exact solution given by Eq. (109) for the infinitely short incident electromagnetic pulse.

The dependence on time of the maximum electric field strength and the maximum local wavenumber in the head of the wavepacket just behind the luminal plasma-vacuum interface is presented in Fig. 25. The former is well approximated as $E_{\max} \propto t^{-0.32}$ while the latter – as $k_{\max} \propto t^{0.65}$. We note that these maxima are determined by the smoothness of the profile function $\rho$ at the interface. The information on how much energy is stored in the maximum wavenumber is seen in Fig. 26 which presents the fast Fourier transform of the function $E(x)$ for different moments of time. As time progresses, the spectrum is extended to higher wavenumbers along the curve $\propto k^{-1}$. We see that the unbounded increase of both the wavenumber and the product of the wavenumber and the electric field strength does not violate the energy conservation. The energy stored in the spectrum is $\propto \int I_k^2 dk = \text{const} + \int_{k_*}^{\infty} k^{-2} dk < \infty$, where $k_*$ indicates the start of the $k^{-1}$ asymptotic in Fig. 26. As also seen in Fig. 26, the energy for higher wavenumbers at the next time moment is apparently taken from the spectral maximum near the cutoff at the previous time moment. The asymptotic behaviour of the spectrum is well predicted by Eq. (110) and illustrated in Fig. 19(a) for the sharp plasma-vacuum interface. We note that Eq. (110) is for the case of an infinitely sharp plasma-vacuum interface, whereas here the interface is smooth with the characteristic wavenumber of $\sim 2\pi/l_{\text{slope}} = 10\pi$; in Fig. 26 much higher wavenumbers are seen, similarly to an infinitely sharp interface.

Fig. 27 shows the electric field strength distribution in the plane $(x, t)$ for the short electromagnetic pulse, Eq. (43), interaction with the luminal plasma slab, Eq. (39). It coincides with Fig. 21 below the world line of the rear plasma-vacuum interface. Above that world line we see a transmission with relatively slowly decreasing magnitude and frequency. The field congestion at the interface is the same as in the case of the infinitely thick plasma layer, as seen in Figs. 28-29. Not surprisingly, the field inside the luminal slab, Fig. 29, as well as its local wavenumber, is identical to the same-length portion just behind the luminal plasma-vacuum interface in Fig. 23 (except only a small portion near the rear side of the slab). Therefore, it has the same asymptotic behaviour near the interface.

The *phase object*, in particular, the luminal plasma slab, can disappear almost instantaneously if an agent causing the local modification of medium properties, in particular, the Langmuir frequency, ceases its action. As an example, one can imagine a focusing searchlight so intense that it induces a change of the refractive index of ambient gas at the moving focus; when it is turned off, the refractive index returns to its normal value.

This rise the question: if the luminal plasma slab disappears (i.e. the local Langmuir frequency vanishes), what will happen to the high frequency radiation stored in the slab?

We preformed simulations, where the plasma slab disappears at some time either abruptly or by a linear-in-time decrease of its profile function with some timescale. In other words, for abrupt disappearance at $t = t_X$:

$$(113) \quad \rho_a = \begin{cases} \text{Eq. (39)} & \text{for } t < t_X; \\ 0 & \text{for } t \geq t_X, \end{cases}$$

and for a gradual decrease starting from $t = t_X$ till $t = t_X + \Delta t_X$:

$$\text{(114)} \quad \rho_g = \begin{cases} \text{Eq. (39)} & \text{for } t < t_X; \\ \left(1 - \frac{t-t_X}{\Delta t_X}\right) \times [\text{Eq. (39)}] & \text{for } t_X \leq t < t_X + \Delta t_X, \\ 0 & \text{for } t_X \geq t_X + \Delta t_X. \end{cases}$$

The results are presented in Figs. 30-31 for $t_X = 400$. We see the release of the electromagnetic wavepacket initially confined in the luminal plasma slab that disappears either instantaneously, Fig. 30, or gradually with $\Delta t_X = 40$, Fig. 31. From the electromagnetic field strength distribution in the $(x - t, t)$ plane, we see the change of the equal-phase curves. Inside the luminal plasma slab, they are inclined; the inclination angle tangent is $1/(\beta_{ph} - 1)$ where $\beta_{ph}$ is the wave phase velocity as in Eq. (80). When the profile function $\rho$ vanishes, the equal-phase curves becomes straight vertical as it should be in vacuum in the $(x - t, t)$ plane. We note that the local wavenumber almost preserved during the luminal plasma slab disappearance. In other words, the frequency of the released electromagnetic wavepacket simply becomes fixed after the slab disappearance.

## 4. Terminology

In previous sections we formally named the electromagnetic waves introduced into and induced during the interaction as "incident", "reflected" or "co-reflected", and "transmitted" waves. We found that due to dispersion effects the "transmitted" wave can have negative or positive phase and group velocity. When the phase and group velocity are positive, the transmitted wave actually becomes a reflected wave. In the case of subluminal plasma-vacuum interface, it can leave plasma. In the case of luminal and superluminal interface, it stays in plasma behind the interface; but it can be released almost "as is" when the refractive index suddenly vanishes. In the case of subluminal and superluminal interface, this "extraordinary" reflection makes a wavepacket in addition to the "ordinary" reflection at the interface (called above "reflection" in the subluminal case or "co-reflection" in the superluminal case). In the case of luminal interface, the "ordinary" reflection disappears completely, while the "extraordinary" reflection due to dispersion effects exists. In the light of this discussion we can answer the question "Can the luminal mirror reflect?" in the following way. The luminal interface does not have ordinary reflection in contrast to the subluminal or superluminal interface; in this sense the reflectivity of the luminal interface is zero. However, due to dispersion effects the luminal interface can reflect extraordinarily. In the case of an incident plane wave extraordinary reflectivity is $a_{R(T)} = 1$, according to Eq. (53), while the extraordinary reflected frequency upshift factor is $\omega_{R(T)} = 1 + n/(4n_{cr}) > 2$, according to Eq. (70). When the incident radiation is a short wavepacket, the luminal interface produces a wavepacket whose maximum frequency increases in time; the energy of the high frequency portion is described by the spectrum in Eq. (110).

## 5. Conclusion

We found that in the case of subluminal, luminal, and superluminal plasma-vacuum interface velocity, the transmitted radiation can have negative, zero or positive group velocity, depending on the incident radiation frequency. The zero group velocity corresponds to a standing transmitted wave which eventually can lead to excitation of Langmuir waves [16,17] and electromagnetic solitons [18]. In the cases of sub- and superluminal interface, there is the "ordinarily" reflected wave (which we call co-reflected in the superluminal case) and, additionally, there can be an "extraordinarily" reflected wave that is a transmitted wave propagating in the same direction as the "ordinarily" reflected wave. For the subluminal interface, the "ordinarily" reflected wave emerges at the interface in vacuum, while the "extraordinarily" reflected wave can leave plasma producing a long relatively weak tail following the "ordinarily" reflected wave, with the same frequencye. For the superluminal interface, both types of waves, "ordinarily" and "extraordinarily" reflected, remain in plasma behind the interface. If the superluminal plasma slab is finite, an incident short pulse produces a short symmetric-in-time co-reflected pulse with the length equal to the width of the superluminal plasma slab and with negatively chirped head and positively chirped tail.

As the velocity of the plasma-vacuum interface tends to the speed of light in vacuum, in the cases of both the sub- and superluminal interface, the reflected radiation frequency grows to infinity while the reflection coefficient vanishes, so that the reflected energy tends to zero. Nevertheless, the reflected electric field strength remains finite and non-zero.

By virtue of the Tikhonov theorem [20] we show that the solution of the model of the electromagnetic wave interaction with the luminal interface is a valid limit of the solution of the more general model embracing all three cases of the interface velocity: subluminal, luminal and superluminal.

We show that a long monochromatic pulse with sufficiently high frequency transmits through the luminal interface with a change of its phase, without "ordinary" reflection, unlike the cases of sub- and superluminal interface. However, for sufficiently low incident frequency there is an "extraordinary" reflection, and the "extraordinarily" reflected wave frequency is inversely proportional to the incident wave frequency. A short pulse with a wide spectrum produces behind the luminal interface a wavepacket whose local wavenumber near the interface increases indefinitely, and also increases the product of the local frequency and electric field strength, in contrast to the cases of the sub- and superluminal interface. We found that if the luminal interface disappears, either instantaneously or gradually, the high-frequency "extraordinarily" reflected radiation is released into vacuum with preserved frequency and magnitude.

The presented results are applicable to media whose refractive index spatial modulations move with subluminal, luminal, or superluminal velocity and can quickly disappear if the agent inducing it ceases its action.

**Acknowledgements**
This work was supported by the NSF and Czech Science Foundation (NSF-GACR collaborative Grant No. 2206059 and NSF Grant No. 2108075) and by the project 'Advanced Research Using High Intensity Laser Produced Photons and Particles' (ADONIS) CZ.02.1.01/0.0/0.0/16_019/0000789 from European Regional Development Fund. The authors are thankful to F. Pegoraro, A. S. Pirozhkov, M. Kando, P. V. Sasorov, and P. Valenta for fruitful discussions.

**Figures**

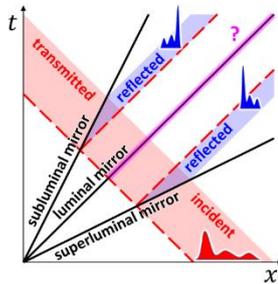

Fig. 1. Minkowski diagram: the worldlines of subluminal, luminal, and superluminal mirrors (black); semi-transparent stripes represent incident and reflected electromagnetic pulses. Dashed lines denote the leading front of the incident pulse which becomes the front or rear of the pulse reflected from the sub- or superluminal mirror, respectively.

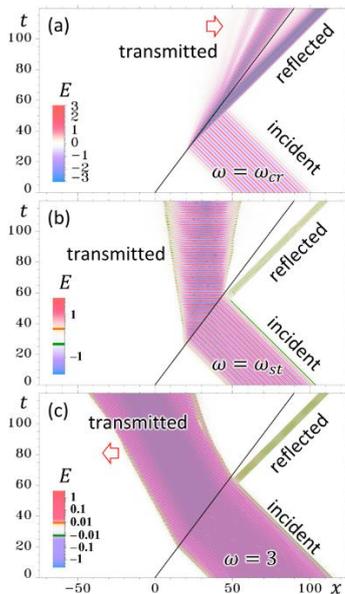

Fig. 2. The subluminal plasma-vacuum interface and the long monochromatic electromagnetic pulse interaction in the $(x, t)$ plane. The interface worldline (black) and the electric field strength (blue-red) are shown for the incident wave frequency of (a) $\omega_{cr}$, (b) $\omega_{st}$, and (c) $\omega = 3$. Hollow arrows show the wave propagation direction. The colorscale for (a) is saturated, i.e. values out of the range are shown by the same color as the corresponding range limits. For (b) and (c) the colorscale is logarithmic, with emphasis (green-orange) on the magnitude of the reflected radiation.

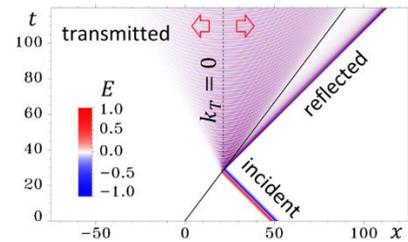

Fig. 3. The subluminal plasma-vacuum interface and the short electromagnetic pulse interaction in the $(x, t)$ plane. Dashed line denotes the location where the wavenumber is formally zero (in the plane wave approximation). Each hollow arrow shows the direction of the wave phase and group velocity. The colorscale is saturated.

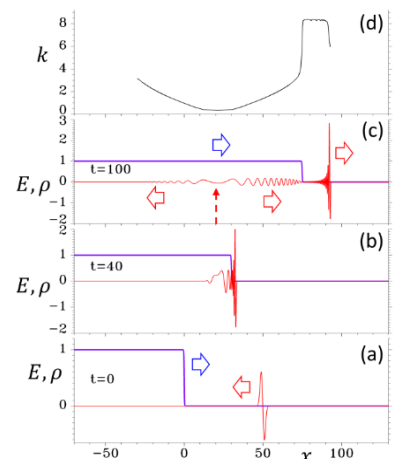

Fig. 4. The cross-sections of Fig. 3 at different moments of time: (a,b,c) the subluminal plasma-vacuum interface profile function $\rho(x)$ and the electric field strength $E(x)$. (d): The local wavenumber $k(x)$ of the electric field strength in (c). Hollow arrows show the direction of propagation of the plasma interface or waves. The vertical dashed arrow shows the location of a formally zero

wavenumber.

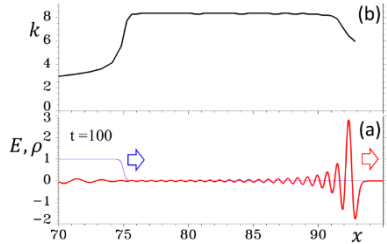

Fig. 5. The reflected pulse (a) and its local wavenumber (b), the closeups the frames (c) and (d) of Fig. 4, respectively.

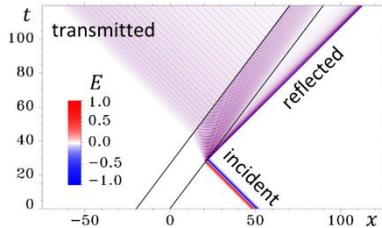

Fig. 6. The subluminal plasma slab and the short electromagnetic pulse interaction in the $(x,t)$ plane shown by the worldlines of the slab front and rear (black lines) and by the electric field strength (blue-red). The colorscale is saturated.

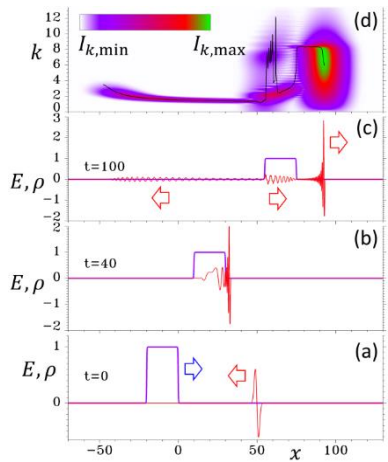

Fig. 7. The cross-sections of Fig. 6 at different moments of time: (a,b,c) the subluminal plasma slab profile function $\rho(x)$ and the electric field strength $E(x)$. (d): The local spectrum intensity in arb.u. (windowed fast Fourier transform of $E(x)$ at every point $x$, see text) with the superimposed local wavenumber $k(x)$ of the electric field strength in (c). Hollow arrows show the propagation direction of the plasma interface or waves.

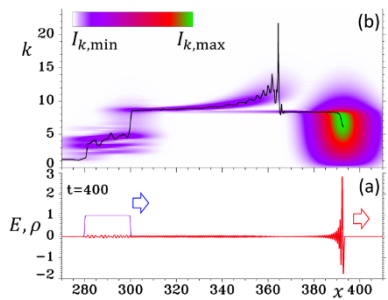

Fig. 8. The same entities as in Fig. 7(c,d) at later time.

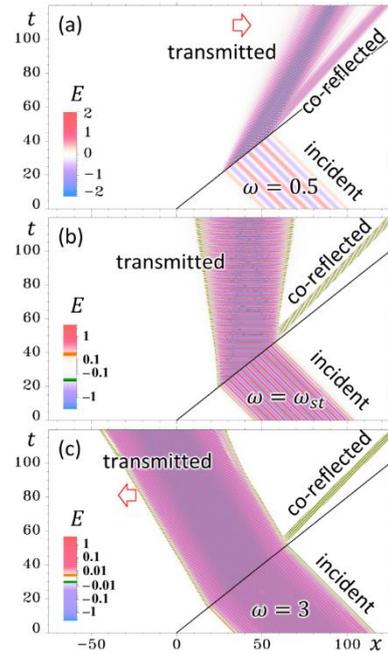

Fig. 9. The superluminal plasma-vacuum interface and the long monochromatic electromagnetic pulse interaction in the $(x,t)$ plane. The interface worldline (black) and the electric field strength (blue-red) are shown for the incident wave frequency of (a) $\omega = 0.5$, (b) $\omega_{st}$, and (c) $\omega = 3$. Hollow arrows show the wave propagation direction. The colorscale for (a) is saturated, for (b) and (c) the colorscale is logarithmic, with emphasis (green-orange) on the magnitude of the reflected radiation.

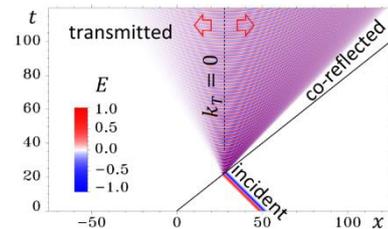

Fig. 10. The superluminal plasma-vacuum interface and the short electromagnetic pulse interaction in the $(x,t)$ plane. Dashed line denotes the location where the wavenumber is formally zero (in the plane wave approximation). Each hollow arrow shows the direction of the wave phase and group velocity. The colorscale is saturated.

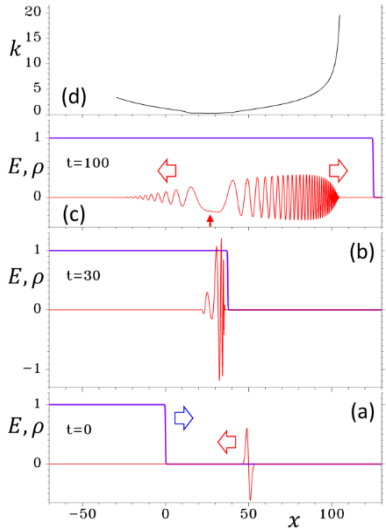

Fig. 11. The cross-sections of Fig. 10 at different moments of time: (a,b,c) the superluminal plasma-vacuum interface profile function $\rho(x)$ and the electric field strength $E(x)$. (d): The local wavenumber $k(x)$ of the electric field strength in (c). Hollow arrows show the direction of propagation of the plasma interface or waves. The vertical arrow in (c) shows the location of a formally zero wavenumber.

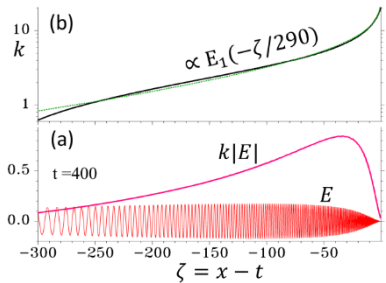

Fig. 12. The electric field strength (a) and its local wave number (b) at later time, showing a longer evolution of entities of Fig. 11(c,d) and the product of the electric field local wavenumber and envelope magnitude $k|E|$. In (a), the ordinate axis is for both the electric field strength $E$ and $k|E|$. In (b), $E_1(\zeta)$ is the exponential integral E of the order 1.

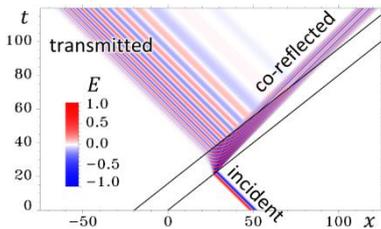

Fig. 13. The superluminal plasma slab and the short electromagnetic pulse interaction in the $(x, t)$ plane shown by the worldlines of the slab front and rear (black lines) and by the electric field strength (blue-red). The colorscale is saturated.

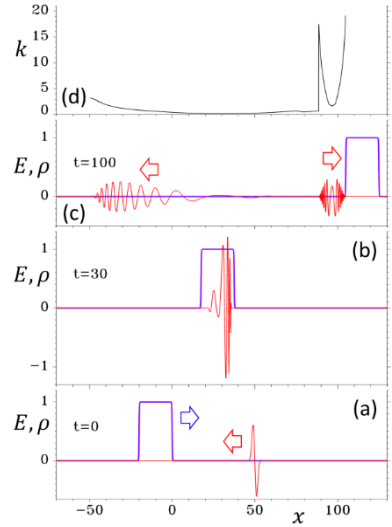

Fig. 14. The cross-sections of Fig. 13 at different moments of time: (a,b,c) the superluminal plasma slab profile function $\rho(x)$ and the electric field strength $E(x)$. (d): The local wavenumber $k(x)$ of the electric field strength in (c). Hollow arrows show the propagation direction of the plasma interface or waves.

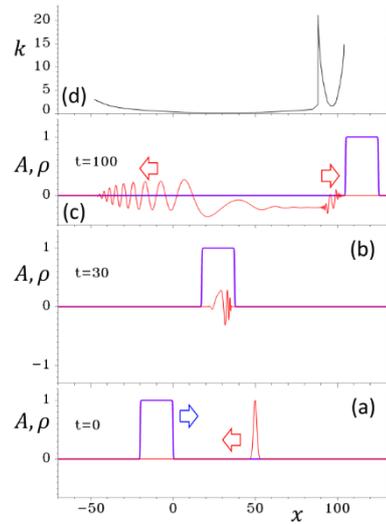

Fig. 15. The evolution of the vector-potential $A(x)$ corresponding to the electric field strength $E(x) = -\partial_t A$ of Fig. 14. (a,b,c) The superluminal plasma slab profile function $\rho(x)$ and the vector-potential $A(x)$. (d): The local wavenumber $k(x)$ of the vector-potential in (c). Hollow arrows show the propagation direction of the plasma interface or waves.

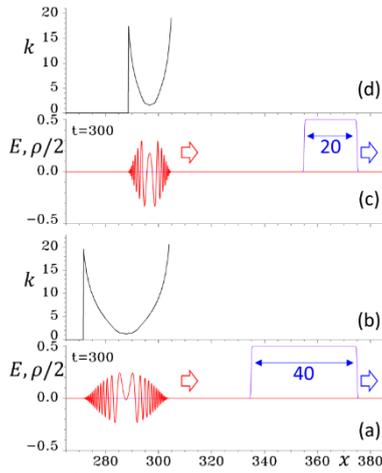

Fig. 16. Co-reflected electromagnetic pulse behind the superluminal plasma slab of the width of 20 (c,d) and 40 (a,b). (a,c) The profile function $\rho(x)$ and the electric field strength $E(x)$. (b,d) The local wave number of $E(x)$ corresponding, respectively, to (a,c).

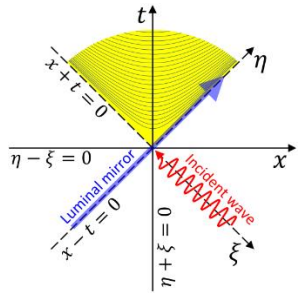

Fig. 17. The $(x,t)$ and $(\xi,\eta)$ variables, worldlines of the luminal mirror and incident wave, and the light cone (region of influence) of the event of the incident wave and luminal mirror collision.

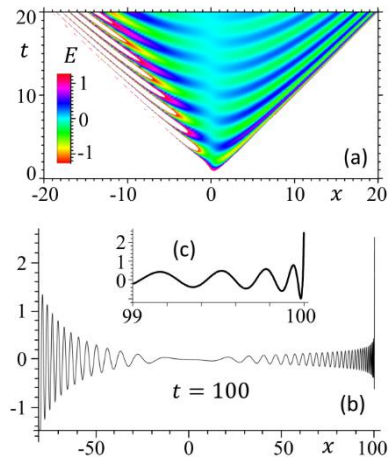

Fig. 18. The electric field strength according to Eq. (109): (a) the distribution in the $(x,t)$ plane; (b) the dependence on $x$ at $t = 100$; (c) the behaviour near $x = t$; the limit is $E(x = t) = \sqrt{2\pi}$.

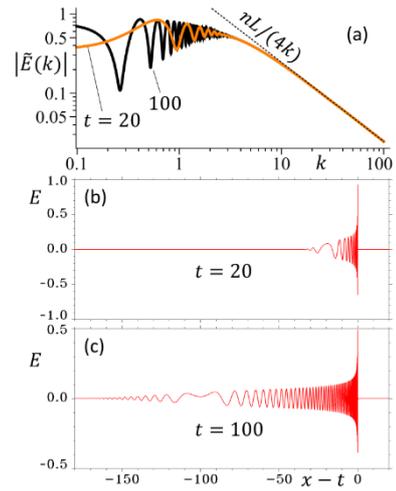

Fig. 19. The modulus of the right-hand side of Eq. (110) at different time (a) and the corresponding wavepackets (b,c) in terms of the electric field strength (in arbitrary units).

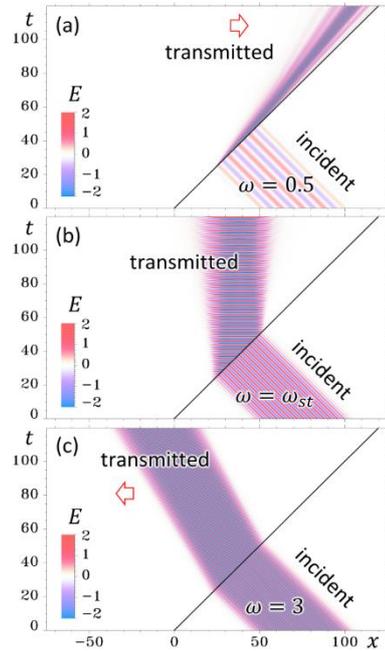

Fig. 20. The luminal plasma-vacuum interface and the long monochromatic electromagnetic pulse interaction in the $(x,t)$ plane. The interface worldline (black) and the electric field strength (blue-red) are shown for the incident wave frequency of (a) $\omega = 0.5$, (b) $\omega_{st}$, and (c) $\omega = 3$. Hollow arrows show the wave propagation direction. The colorscale is saturated and is the same for (a,b,c).

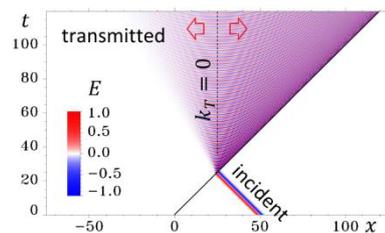

Fig. 21. The luminal plasma-vacuum interface and the short electromagnetic pulse interaction in the $(x,t)$ plane. Dashed line denotes the location

where the wavenumber is formally zero (in the plane wave approximation). Each hollow arrow shows the direction of the wave phase and group velocity. The colorscale is saturated.

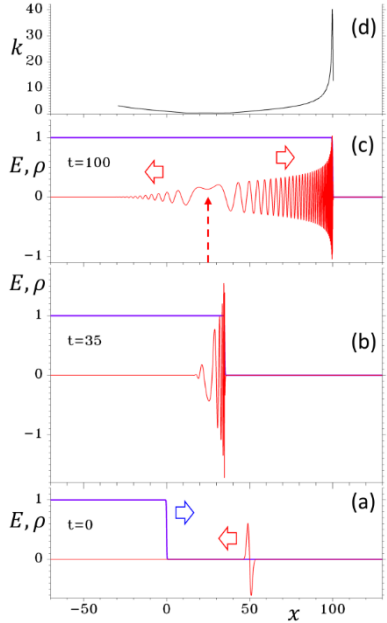

Fig. 22. The cross-sections of Fig. 21 at different moments of time: (a,b,c) the luminal plasma-vacuum interface profile function $\rho(x)$ and the electric field strength $E(x)$. (d): The local wavenumber $k(x)$ of the electric field strength in (c). Hollow arrows show the direction of propagation of the plasma interface or waves. The vertical dashed arrow shows the location of a formally zero wavenumber.

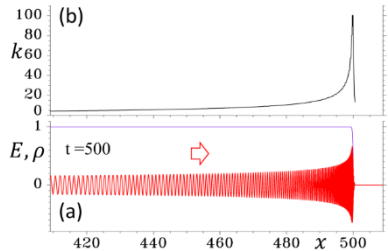

Fig. 23. The same entities as in Fig. 22(c,d) at later time: the electric field strength (a) and its local wave number (b). Note the increase of local wave number near the luminal plasma-vacuum interface compared to Fig. 22(d).

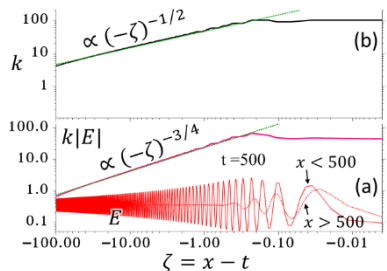

Fig. 24. The same entities as in Fig. 23 and $k|E|$, the product of the local wavenumber $k$ and envelope magnitude $|E|$ of the electric field strength in a log-log scale. In (a), the ordinate axis in logarithmic scale is only for $k|E|$, the abscissa axis in logarithmic scale is for both $k|E|$ and $E$. For the electric field strength $E$ the ordinate axis (not shown) is linearly scaled; the solid and dashed curves are for $x < 500$ and $x > 500$, respectively. The local wavenumber $k$ and the product $k|E|$ are shown only for $x < 500$.

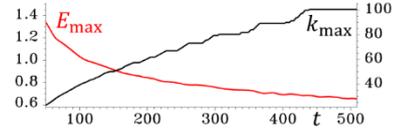

Fig. 25. The dependence on time of the maximum electric field strength $E_{max}$ (left axis) and the maximum local wavenumber $k_{max}$ (right axis) of the wavepacket behind the luminal interface, seen in simulations. Approximately, $E_{max} \propto t^{-0.32}$ and $k_{max} \propto t^{0.65}$.

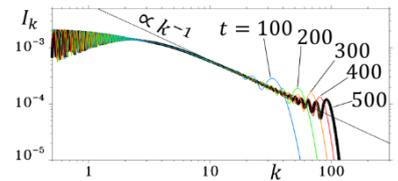

Fig. 26. The spatial spectrum of the electric field strength $E$ at different time moments, as seen in simulations. $I_k$ is the absolute value of the fast Fourier transform of $E(x)$.

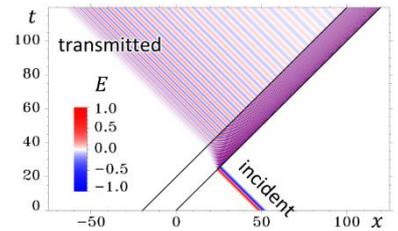

Fig. 27. The luminal plasma slab and the short electromagnetic pulse interaction in the $(x, t)$ plane shown by the worldlines of the slab front and rear (black lines) and by the electric field strength (blue-red). The colorscale is saturated.

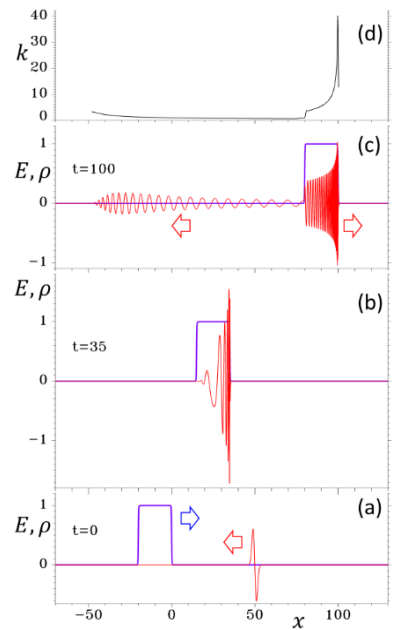

Fig. 28. The cross-sections of Fig. 27 at different moments of time: (a,b,c) the luminal plasma slab profile function $\rho(x)$ and the electric field strength $E(x)$. (d): The local wavenumber $k(x)$ of the electric field strength in (c). Hollow arrows show the propagation direction of the plasma interface or waves.

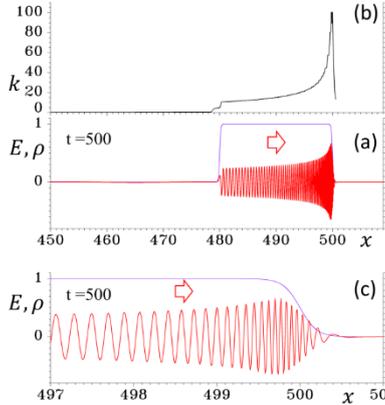

Fig. 29. The same entities as in Fig. 28(c,d) at later time: the electric field strength (a) and its local wave number (b). Note the increase of local wave number near the front of the luminal plasma slab compared to Fig. 28(d). (c): The closeup of (a) near the front of the luminal plasma slab.

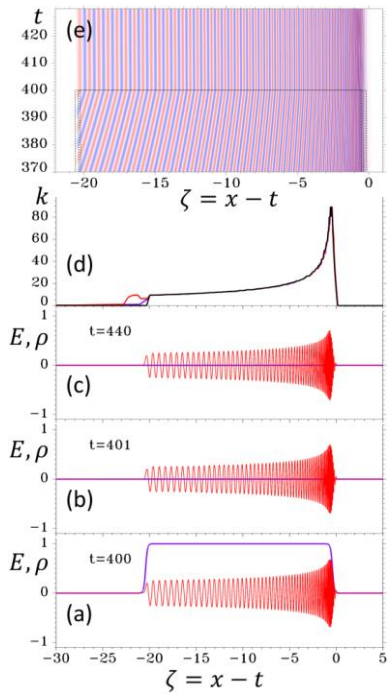

Fig. 30. The release of the electromagnetic wavepacket initially confined in the luminal plasma slab that instantaneously disappears at $t = t_X = 400$ according to Eq. (113). (e) The electric field strength distribution and the isocurves of the profile function $\rho(x, t)$ for 0.8, 0.5, 0.1 in the $(x - t, t)$ plane. (a,b,c) The luminal plasma slab profile function $\rho(x)$ and the electric field strength $E(x)$ at different time moments. (d): The local wavenumber $k(x)$ of the electric field strength in (a,b,c) marked by (blue,red,black), respectively.

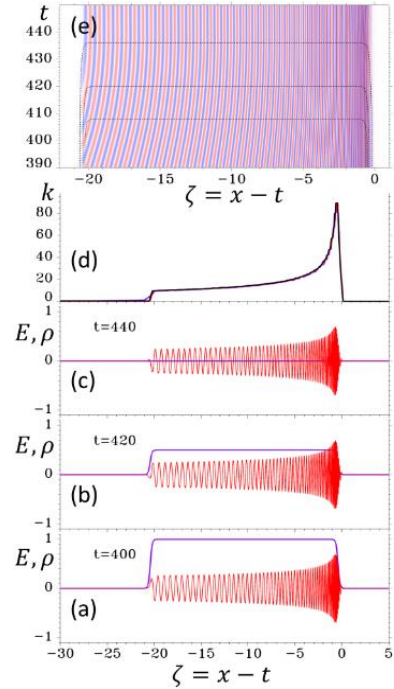

Fig. 31. The same entities as in Fig. 30 for the luminal plasma slab that disappears gradually, from $t = 400$ till $t = 440$, according to Eq.(114).

**References**

[1] A. Einstein, Zur Elektrodynamik bewegter Koerper, Ann. Phys. (Leipzig) 17, 891 (1905).
[2] B. M. Bolotovskii, V. L. Ginzburg, The Vavilov–Cerenkov effect and the Doppler effect in the motion of sources with superluminal velocity in vacuum, Physics Uspekhi, 106, 577 (1972).
[3] R. I. Sutherland and J. R. Shepanski, Superluminal reference frames and generalized Lorentz transformations, Phys. Rev. D 33, 2896 (1986).
[4] T. M. Jeong, S. V. Bulanov, P. Hadjisolomou, T. M. Esirkepov, Superluminal-subluminal orbital angular momentum femtosecond laser focus, Optics Express 29, 31665 (2021).

[5] T. Z. Esirkepov, S. V. Bulanov, M. Kando, A. S. Pirozhkov, and A. G. Zhidkov, Boosted high-harmonics pulse from a double-sided relativistic mirror, Phys. Rev. Lett. 103, 025002 (2009).

[6] T. Zh. Esirkepov, S. V. Bulanov, A. G. Zhidkov, A. S. Pirozhkov and M. Kando, High-power laser-driven source of ultra-short X-ray and gamma-ray pulses, Eur. Phys. J. D 55, 457 (2009).

[7] A. Zhidkov, T. Esirkepov, T. Fujii, K. Nemoto, J. Koga, S. V. Bulanov, Characteristics of light reflected from a dense ionization wave with a tunable velocity, Phys. Rev. Lett. 103, 215003 (2009).

[8] Z. Bu, B. Shen, S. Huang, S. Li, and H. Zhang, Light reversing and folding based on a superluminal flying mirror in a plasma with increasing density, Plasma Phys. Control. Fusion 58, 075008 (2016).

[9] S. V. Bulanov, T. Zh. Esirkepov, M. Kando, A. S. Pirozhkov, and N. N. Rosanov, Relativistic Mirrors in Plasmas–Novel Results and Perspectives, Physics Uspekhi 56, 429 (2013).

[10] S. V. Bulanov, T. Zh. Esirkepov, T. Tajima, Light intensification towards the Schwinger limit. Phys. Rev. Lett. 91, 085001 (2003).

[11] S. S. Bulanov, A. Maksimchuk, C. B. Schroeder, A. G. Zhidkov, E. Esarey, W. P. Leemans, Relativistic spherical plasma waves. Physics of Plasmas 19, 020704 (2012).

[12] V. V. Kulagin, V.A. Cherepenin, M. S. Hur, H. Suk, Flying mirror model for interaction of a super-intense nonadiabatic laser pulse with a thin plasma layer: Dynamics of electrons in a linearly polarized external field. Phys. Plasmas 14(11) (2007)

[13] S. V. Bulanov, N. M. Naumova, F. Pegoraro, Interaction of an ultrashort, relativistically strong laser pulse with an overdense plasma. Phys. Plasmas 1, 745 (1994).

[14] H. Vincenti, Achieving extreme light intensities using optically curved relativistic plasma mirrors, Physical review letters 123, 105001 (2019).

[15] A. Taflove and S. C. Hagness, Computational Electrodynamics: The Finite-Difference Time-Domain Method, Artech House on Demand; 3rd edition (June 30, 2005).

[16] M. N. Rosenbluth and C. S. Liu, Excitation of Plasma Waves by Two Laser Beams, Phys. Rev. Lett. 29, 701 (1972).

[17] T. Tajima and J. M. Dawson, Laser Electron Accelerator, Phys. Rev. Lett. 43, 267 (1979).

[18] V. A. Kozlov, A. G. Litvak, E. V. Suvorov, Envelope solitons of relativistic strong electromagnetic waves, JETP, 49, 75 (1979).

[19] V. L. Ginzburg, The Propagation of Electromagnetic Waves in Plasmas (Pergamon, New York, 64).

[20] A. N. Tikhonov, Systems of differential equations containing small parameters in the derivatives, Mat. Sb. (N.S.), 73, 575 (1952).

[21] I. S. Gradshteyn, I. M. Ryzhik, Yu. V. Geronimus, M. Yu. Tseytlin (February 2007). A. Jeffrey, D. Zwillinger (eds.). Table of Integrals, Series, and Products. Translated by Scripta Technica, Inc. (7 ed.). Academic Press, Inc. Entry 3.697.